\documentstyle[floats,prd,aps,epsf]{revtex}
\draft 

\begin{document}


\twocolumn[\hsize\textwidth\columnwidth\hsize\csname
@twocolumnfalse\endcsname

\title{The Innermost Stable Circular Orbit of Binary Black Holes}

\author{Thomas W.~Baumgarte}

\address{Department of Physics, University of Illinois at
	Urbana-Champaign, Urbana, Il~61801}

\maketitle

\begin{abstract}
We introduce a new method to construct solutions to the constraint
equations of general relativity describing binary black holes in
quasicircular orbit.  Black hole pairs with arbitrary momenta can be
constructed with a simple method recently suggested by Brandt and
Br\"ugmann, and quasicircular orbits can then be found by locating a
minimum in the binding energy along sequences of constant horizon
area.  This approach produces binary black holes in a "three-sheeted"
manifold structure, as opposed to the "two-sheeted" structure in the
conformal-imaging approach adopted earlier by Cook.  We focus on
locating the innermost stable circular orbit and compare with earlier
calculations.  Our results confirm those of Cook and imply that the
underlying manifold structure has a very small effect on the location
of the innermost stable circular orbit.
\end{abstract}


\vskip2pc]


\section{INTRODUCTION}

Binary black holes are among the most promising sources of
gravitational radiation for the new generation of gravitational wave
detectors such as the Laser Interferometric Gravitational Wave
Observatory (LIGO), VIRGO, GEO and TAMA.  This has motivated an
intense theoretical effort to predict the gravitational waveform
emitted during the inspiral and coalescence of two black
holes~\cite{itp00}.

Because of the circularizing effects of gravitational radiation
damping, we expect the orbits of close binary systems to have small
eccentricities.  The inspiral of a binary black hole system then
proceeds adiabatically along a sequence of quasicircular orbits up to
the innermost stable circular orbit (hereafter ISCO), where the
evolution is expected to change into a rapid plunge and 
coalescence\cite{footnote0}.
The ISCO therefore leaves a characteristic signature in the
gravitational wave signal, and knowledge of its location and frequency
is thus very important for the prospect of future observations.

While various approximations may be adequate to model the adiabatic
inspiral up to the ISCO, it is generally expected that only numerical
simulations in full general relativity can accurately model the
dynamical plunge and merger and predict the gravitational signal from
that phase.  It is therefore desirable to construct initial data for
numerical evolution calculations describing binary black hole pairs at
the ISCO, which adds another motivation for determining the 
location of the ISCO.

Various approaches have been adopted to locate the ISCO in compact
binaries, including first order post-Newtonian
approximations~\cite{ce77}, variational principles~\cite{bd92}, second
order post-Newtonian methods combined with a ``hybrid''
approach~\cite{kww92}, a Pad\'{e} approximation~\cite{dis98} and an
effective-one-body approach~\cite{bd99,bd00}, and numerical solutions
to the constraint equations of general relativity~\cite{c94,betal98}.
Unfortunately, however, the results differ significantly and yet have
to show any sign of convergence (see Table~II below).  It would
clearly be desirable to understand the origin of these differences.
In this paper, we revisit binary black hole solutions to the
constraint equations, and evaluate how some of the choices which have
to be made in this approach affect the location of the ISCO.

Before the constraint equations of general relativity can be solved, a
background geometry and topology have to be chosen.  In the
conformal-imaging approach adopted by Cook~\cite{c94}, a conformally
flat (spatial) background metric is chosen together with a two-sheeted
manifold structure (see Sec.~\ref{sec2.1}).  It has been suggested
that these choices may affect the location of the ISCO, and may
explain the difference between these and the more recent post-Newtonian
results.

In this paper, we combine the methods of Cook~\cite{c94} and Brandt
and Br\"ugmann~\cite{bb97} to introduce a new approach to constructing
binary black holes in quasicircular orbit.  We follow Cook~\cite{c94}
and choose a conformally flat background metric, but do not assume an
inversion-symmetry as is done in the conformal-imaging approach.  This
considerably simplifies the solution of the momentum constraint (see
Sec.~\ref{sec2.2}), and produces binary black holes in a three-sheeted
manifold structure as opposed to the two-sheeted structure in the
conformal-imaging approach.  Moreover, adopting the ``puncture''
approach of Brandt and Br\"ugmann~\cite{bb97}, the Hamiltonian
constraint can be solved very easily numerically on R$^3$ without
having to impose boundary conditions on interior boundaries (see
Sec.~\ref{sec2.3}).  We locate the ISCO, and find that its physical
parameters agree very well with those found with the conformal-imaging
approach of Cook~\cite{c94}.  We therefore conclude that the choice of
the underlying mani\-fold structure has a very small effect on the
location of the ISCO.  Our new approach, which is significantly
simpler than the conformal-imaging approach, may also provide a
framework in which the conformal-flatness assumption may be relaxed,
and its effect on the ISCO be evaluated.

The paper is organized as follows.  In Sec.~II, we introduce the basic
equations and explain how binary black holes in quasicircular orbit
can be constructed.  We discuss our numerical implementation in
Sec.~III.  In Sec.~IV we present our results and compare with those from
other approaches.  We briefly summarize in Sec.~V.


\section{SETUP OF THE PROBLEM}

\subsection{The Initial Value Problem}
\label{sec2.1}

A framework for constructing initial data describing binary black
holes has been provided by Arnowitt, Deser and Misner's $3+1$
decomposition of Einstein's equations~\cite{adm62} and York's
conformal decomposition~\cite{y71,y79}.  

The $3+1$ decomposition splits Einstein's equations into evolution and
constraint equations for the metric $\gamma_{ij}$ of a
spatial hypersurface $\Sigma$, and the extrinsic curvature $K_{ij}$,
which describes the embedding of the hypersurface $\Sigma$ in the full
spacetime.  The physical metric $\gamma_{ij}$ can now be decomposed
into a conformal factor $\psi$ and a conformal background metric 
$\hat \gamma_{ij}$,
\begin{equation}
\gamma_{ij} = \psi^4 \hat \gamma_{ij}.
\end{equation}
It is also convenient to decompose the extrinsic curvature $K_{ij}$ into
its trace $K$ and a trace-free conformal background extrinsic curvature 
$\hat A_{ij}$ according to
\begin{equation}
K_{ij} = \psi^{-2} \hat A_{ij} + \frac{1}{3} \gamma_{ij} K.
\end{equation}
The Hamiltonian constraint then reduces to an
equation for the conformal factor $\psi$,
\begin{equation}
8 \hat \nabla^2 \psi - \psi \hat R - \frac{2}{3} \psi^5 K^2
	+ \psi^{-7} \hat A_{ij} \hat A^{ij} = 0,
\end{equation}
and the momentum constraint can be written 
\begin{equation}
\hat D_j \hat A^{ij} - \frac{2}{3} \psi^{6} \hat \gamma^{ij} \hat
D_j K = 0.
\end{equation}
Here $\hat D_i$ is the covariant derivative compatible with the
conformal background metric, $\hat \nabla^2$ the Laplacian, and $\hat
R$ is the Ricci scalar.

Binary black hole initial data cannot be constructed uniquely, because
the constraint equations of general relativity determine neither the
background {\em geometry} nor the {\em topology} of the spacetime,
both of which have to be chosen before the constraint equations can be
solved.  Loosely speaking, these ambiguities correspond to different
amounts of gravitational radiation in the initial data sets.  In
this paper, we will follow Cook {\it et.al.}~\cite{c94,c91,cetal93} and
choose a flat background geometry, but we will choose a three-sheeted
topology as opposed to the two-sheeted topology of Cook.

Choosing the background {\em geometry} amounts to choosing the
conformal background metric $\hat \gamma_{ij}$.  Following Cook {\it
et.al.}~\cite{c94,c91,cetal93}, we choose the conformal background
geometry to be flat so that $\hat \gamma_{ij} = f_{ij}$, where $f_{ij}$
is the flat metric in a so far arbitrary coordinate system.  The
covariant derivative $\hat D_i$ then becomes the flat-space covariant
derivative, and the Ricci scalar $\hat R$ vanishes.  We will later
specialize to cartesian coordinates, $\hat \gamma_{ij} = \delta_{ij}$,
for which $\hat D_i$ reduces to a partial derivative.  We also take
the hypersurface $\Sigma$ to be maximally embedded in the spacetime so
that $K = 0$.  With these choices, the constraint equations simplify
to
\begin{equation} \label{ham2}
\hat \nabla^2 \psi 
	= - \frac{1}{8} \psi^{-7} \hat A_{ij} \hat A^{ij}
\end{equation}
and
\begin{equation} \label{mom2}
\hat D_j \hat A^{ij} = 0.
\end{equation}
Note that maximal slicing $K = 0$ automatically decouples the momentum
constraint from the Hamiltonian constraint.

Choosing the {\em topology} of the spacetime is less straight-forward
(compare the discussion in~\cite{c91}).  Since we are interested in
isolated black-holes systems, it is natural to assume the hypersurface
$\Sigma$ to be asymptotically flat.  Constructing black hole data in
vacuum, however, necessarily involves non-trivial topologies.  This
can be illustrated by a $t=const$ slice of the Schwarzschild
geometry in isotropic coordinates, where every point inside the black
hole's throat can be mapped into a point outside the throat and vice
versa.  Moreover, such a mapping can be accomplished with an isometry,
which maps the metric into itself, implying that the physical fields
at a point inside the throat are identical to those at a point outside
the throat.  In particular, the geometry near the center is identical
to the geometry near infinity.  We can therefore think of this
solution as describing two identical, asymptotically flat
``universes'' or ``sheets'', which are connected by a throat or
Einstein-Rosen bridge~\cite{er35}.  

There is no unique generalization of this topology to the case of
multiple black holes~\cite{l63}.  For two black holes, the two throats
could either connect to the same asymptotically flat sheet, or else to
two separate asymptotically flat sheets.  The former approach results
in a two-sheeted topology, the latter in a three-sheeted topology.

Cook {\it et.al.}~\cite{c94,c91,cetal93} implemented a
``con\-for\-mal-im\-ag\-ing'' formalism, which adopts a two-sheeted
topology together with the additional demand that the two sheets are
related by an isometry so that their physical fields are identical
({\it cf.}~\cite{m63,ksy83}).  It has been argued that this choice is
the ``most faithful generalization of the Schwarzschild geometry to
the case of multiple holes''~\cite{c91}.  Moreover, the isometry
conditions on the throats can be used as boundary conditions in
numerical implementations, so that singularities inside the throats
can be eliminated from the numerical grid.  The computational
disadvantage of this method is that boundary conditions have to be
imposed on fairly complicated surfaces.  In finite difference
algorithms, this can be accomplished either with bispherical or
\v{C}ade\v{z} coordinates~\cite{c71}, designed such that a constant
coordinate surface coincides with the throat, or else with fairly
complicated algorithms in cartesian coordinates.  Both approaches,
together with a spectral method, have been compared in~\cite{cetal93}.

In this paper, we choose instead a three-sheeted topology and do not
assume an inversion-symmetry across the throats, which simplifies the
problem in two respects.  The analytical solution to the momentum
constraint becomes very simple, since we no longer need to construct
inversion-symmetric solutions (see Sec.~\ref{sec2.2}).  Moreover, the
singularities inside the black holes can be removed analytically using
a ``puncture'' method recently suggested by Brandt and
Br\"ugmann~\cite{bb97} (see Sec.~\ref{sec2.3}).  The problem can then
be solved quite easily on R$^3$ in cartesian coordinates, without
having to impose interior boundary conditions.  The only added
complication is that one now has to locate apparent horizons in the
numerically constructed hypersurface.

\subsection{Solving the Momentum Constraint for Binary Black Holes}
\label{sec2.2}

For maximally sliced hypersurfaces, the momentum constraint decouples
from the conformal factor, and analytical solutions to~(\ref{mom2})
can be given.  Moreover, for conformally and asymptotically flat data,
the total (physical) linear momentum~\cite{footnote2}
\begin{equation} \label{linmom}
P^i = \frac{1}{8\pi} \oint_{\infty} \hat A^{ij} d^2 S_j
\end{equation}
and the total (physical) angular momentum
\begin{equation} \label{angmom}
J_i = \frac{\epsilon_{ijk}}{8 \pi} \oint_{\infty} x^j \hat A^{kl} d^2 S_l
\end{equation}
can be determined from $\hat A^{ij}$ without having to solve the 
Hamiltonian constraint~(\ref{ham2}) (see~\cite{by80}).  

Analytical solutions to the momentum constraint~(\ref{mom2}) describing
single boosted or spinning black holes have been given by Bowen and
York~\cite{by80,b79,y89}.  A solution $\hat A^{ij}$, describing
a single black hole at the coordinate location ${\bf C}$ with linear
momentum ${\bf P}$ is given by
\begin{equation} \label{A_ij}
\hat A^{ij}_{\bf CP} = \frac{3}{2r^2_{\bf C}} 
	\left(P^i n^j_{\bf C} + P^jn^i_{\bf C} +
	(f^{ij} + n^i_{\bf C} n^j_{\bf C}) P_k n^k_{\bf C} \right).  
\end{equation}
Here $r_{\bf C} = \|x^i - C^i\|$ is the coordinate distance to the
center of the black hole and $n^i_{\bf C} = (x^i - C^i)/r$ is the
normal vector pointing away from that center.  Additional terms have
to be added in the conformal-imagine approach for an isometry
condition to hold across the throat.  Note that we have only included
linear momentum terms in this expression, and that we are therefore
restricting our analysis to non-spinning black holes.

Since the momentum constraint~(\ref{mom2}) is linear, we can construct
binary black hole solutions by superposition of single solutions
\begin{equation} \label{A_ij2}
\hat A^{ij} = \hat A^{ij}_{{\bf C}_1{\bf P}_1} 
	+     \hat A^{ij}_{{\bf C}_2{\bf P}_2}.
\end{equation}
From~(\ref{linmom}) and~(\ref{angmom}) we find the that total momentum
of this solution is ${\bf P} = {\bf P}_1 + {\bf P}_2$, and the
angular momentum about the origin of the coordinate system
\begin{equation} \label{J}
{\bf J}  =  {\bf C}_1{\bf \times P}_1 + {\bf C}_2 {\bf \times P}_2.
\end{equation} 

Note that constructing inversion-symmetric solutions for multiple
black holes in the conformal-imaging approach is fairly complicated.
There, the components of the extrinsic curvature are expressed in terms
of an infinite series of recursively defined quantities
(see~\cite{ksy83} and Appendix~A of~\cite{c94}).  Relaxing the
inversion symmetry, so that the extrinsic curvature can be written as
a simple superposition of two solutions~(\ref{A_ij}), therefore
greatly simplifies the problem.

\subsection{Solving the Hamiltonian Constraint for Binary Black Holes}
\label{sec2.3}

Solutions to the Hamiltonian constraint~(\ref{ham2}) can be
constructed by generalizing the Schwarzschild solution in isotropic
coordinates for a static ({\it i.e.} $\hat A_{ij} = 0$) and
spherically symmetric black hole at coordinate location ${\bf C}$,
\begin{equation}
\psi = 1 + \frac{{\cal M}}{2r_{\bf C}}
\end{equation}
(note that asymptotic flatness demands $\psi \rightarrow 1$ as $r
\rightarrow \infty$).  Solutions describing multiple static black
holes can be constructed by adding contributions ${\cal M}/(2r_{\bf
C})$ for each black hole.  To establish an inversion-symmetry,
additional terms would again have to be added~\cite{m63}.

In the ``puncture'' method suggested by Brandt and
Br\"ugmann~\cite{bb97}, a general nonstatic solution to the
Hamiltonian constraint is written as a sum of the static, analytic
contribution plus a term correcting for finite $\hat A_{ij}$.
Adopting their notation, we write
\begin{equation}
\psi = u + \frac{1}{\alpha},
\end{equation}
where $\alpha$ is defined by
\begin{equation}
\frac{1}{\alpha} 
= \frac{{\cal M}_1}{2r_{{\bf C}_1}} + \frac{{\cal M}_2}{2r_{{\bf C}_2}}.
\end{equation}
The Hamiltonian constraint then becomes an equation for the
correction term $u$
\begin{equation} \label{ham3}
\hat \nabla^2 u = - \beta\,(1 + \alpha u)^{-7},
\end{equation}
where we have abbreviated
\begin{equation}
\beta = \frac{1}{8}\, \alpha^7 \, K_{ij}K^{ij}.
\end{equation}
For asymptotic flatness, we impose a Robin boundary condition
$\partial (r\,(u - 1))/\partial r = 0$ at large distances from the
black holes.  The existence and uniqueness of solutions $u$ on R$^3$
has been established in~\cite{bb97}.  The beauty of this approach is
that the poles at the center of the black holes have been absorbed
into the analytical terms. The corrections $u$ are regular everywhere
and can be solved for very easily on a simple computational domain,
without having to impose boundary conditions on the throats.

Once the conformal factor $\psi$ has been determined, the ADM mass of
the solution can be found from
\begin{eqnarray} \label{E}
E & = & - \frac{1}{2 \pi} \oint_{\infty} \hat \nabla^i \psi \,d^2S_i 
	\nonumber \\
& = & - \frac{1}{2 \pi} \oint_{\infty} \hat \nabla^i 
	\left( \frac{1}{\alpha} \right) d^2S_i 
	- \frac{1}{2 \pi} \int \hat \nabla^2 u \,dV \nonumber  \\
& = & {\cal M}_1 + {\cal M}_2 
	+ \frac{1}{2 \pi} \int \beta\,(1 + \alpha u)^{-7} dV.
\end{eqnarray}
Note that the integral extends over all space.

\subsection{Constructing Equal-mass Binary Black Holes in Quasicircular Orbit}

We now specialize to equal mass black holes with ${\cal M} \equiv
{\cal M}_1 = {\cal M}_2$.  In the center-of-mass frame of the binary
system, we have
\begin{equation}
{\bf P} \equiv {\bf P}_1 = - {\bf P}_2.  
\end{equation}
Binaries in quasicircular orbit should furthermore satisfy ${\bf P
\cdot C} = 0$, where we have defined
\begin{equation}
{\bf C} \equiv {\bf C}_1 - {\bf C}_2.
\end{equation}
Without loss of generality, we can then take ${\bf C}$ to be aligned
with the $z$-axis, ${\bf P}$ to be aligned with the $x$-axis, and
place the origin of the cartesian coordinate system at the center
between the two black holes.

The problem has now been reduced to a three-di\-men\-sio\-nal
parameter space with the free parameters ${\cal M}$, $C \equiv \| {\bf
C} \|$ and $P \equiv \| {\bf P } \|$.  For every configuration, we
compute several physical quantities.  We determine the total ADM mass
$E$ from~(\ref{E}) and the total angular momentum $J \equiv J_y = PC$
from~(\ref{J}).  Since we have restricted our analysis to non-spinning
black holes, the mass of each individual black hole can be identified
with the irreducible mass
\begin{equation}
M = M_{\rm irr} \approx \left( \frac{A}{16 \pi} \right)^{1/2},
\end{equation}
where $A$ the proper area of the black hole's apparent
horizon~\cite{c70}.  We now define the effective potential as the binding
energy
\begin{equation} \label{E_b}
E_b = E - 2M.
\end{equation}
Lastly, we compute the proper separation $l$ between the two horizons
along the line connecting the centers of the two apparent horizons,
which is a very good approximation to the shortest proper separation
between the two horizons.

Quasicircular orbits can then be found quite easily (see~\cite{c94})
by computing the effective potential $E_b$ as a function of separation
$l$ along a sequence of constant black hole mass $M$ and angular
momentum $J$ and locating turning points
\begin{equation}
\left. \frac{\partial E_b}{\partial l} \right|_{M,J} = 0.
\end{equation}
A minimum corresponds to a stable quasicircular orbit, while a maximum
corresponds to an unstable orbit.  For a quasicircular orbit, the
binary's orbital angular velocity $\Omega$ as measured at infinity can
then be determined from
\begin{equation}
\Omega = \left. \frac{\partial E_b}{\partial J} \right|_{M,l}
\end{equation}
(see footnote~\cite{footnote3} for a Newtonian illustration).


\section{NUMERICAL IMPLEMENTATION}

We adopt a finite difference approach to solve eq.~(\ref{ham3}) in
cartesian coordinates.  The numerical code is implemented in a
parallel, distributed memory environment using DAGH
software~\cite{dagh}, and the Laplace operator in~(\ref{ham3}) is
inverted using PETSc software~\cite{petsc}.  We linearize
eq.~(\ref{ham3}) and iteratively solve for corrections to approximate
solutions until convergence to a desired accuracy has been achieved.
Since the components of the extrinsic curvature~(\ref{A_ij2}) are
either symmetric or antisymmetric across the coordinate planes $x =
0$, $y =0$ and $z = 0$, it is sufficient to solve the Hamiltonian
constraint in only one octant.

In addition to verifying second order convergence of our code, we have
performed tests in the linear regime by comparing with the linear
analytic solution for black holes boosted towards each other (eq.~(16)
in~\cite{bb97}), and in the nonlinear regime by comparing with the
``A2B8'' dataset~\cite{cetal93}, for which values of the ADM mass in a
three-sheeted manifold structure have been given in Table~I
of~\cite{bb97}.  Note, however, that those masses have erroneously
been calculated for black hole spins with signs opposite to those
given in that table, ``${\bf S}_{1,2} = - {\bf S}_{1,2}$''~\cite{b00}.

Given a solution $\psi$ for a set of parameters ${\cal M}$, $C$ and
$P$, we can locate an apparent horizon and determine the black hole
mass $M$ from the horizon's proper area using the algorithm described
in~\cite{betal96}.  This algorithm expresses the location of a closed
surface in terms of symmetric trace-free tensors, and varies the
expansion coefficients until an outer-most trapped surface has been
found.  We found that for the horizons in this problem, which are
fairly spherical and only very mildly deformed, an expansion up to
order $\ell = 4$ is adequate.  

Before constructing sequences of constant black hole mass $M$, it is
convenient to rescale all variables with respect to that desired value
of the black hole mass $M$.  We introduce the sum of the black hole
masses
\begin{equation}
m \equiv M_1 + M_2 = 2 M
\end{equation}
and the reduced mass
\begin{equation}
\mu \equiv \frac{M_1 M_2}{M_1 + M_2} = \frac{M}{2},
\end{equation}
and define the dimensionless parameters $\bar {\cal M} \equiv {\cal
M}/m$ $\bar C \equiv C/m$ and $\bar P \equiv P/\mu$.  We also rescale
the angular momentum and angular velocity as $\bar J \equiv J/\mu m$
and $\bar \Omega \equiv m \Omega$, and identify the dimensionless
effective potential with the rescaled binding energy $\bar E_b \equiv
E_b/\mu$.

Sequences of constant $\bar J$ can now be constructed by setting $\bar
P = \bar J / \bar C$ for a set of different values of $\bar C$, and by
iterating over $\bar {\cal M}$ until the (dimensionless) numerical
value of the black hole mass $\bar M \equiv M_{\rm num}/M$ has
converged to unity within a desired accuracy for each $\bar
C$~\cite{footnote4}.

For typical cases of interest, the difference between the ADM mass $E$
and the sum of the black hole masses $m$ is quite small.  According
to~(\ref{E_b}), the binding energy is therefore much smaller than
those masses, and its relative numerical error much larger.  In order
to reliably locate a minimum in the binding energy, we therefore have
to determine the masses to very high accuracy ({\it cf.}~\cite{c94}).
Moreover, numerous models have to be calculated to construct a
sufficient number of sequences over a sufficient range of separations,
which is only affordable for a fairly moderate maximum grid size.  For
a uniform, cartesian grid of a given size, a compromise then has to be
found between extending the computational grid to large enough
distances and sufficiently resolving the individual black holes.

\begin{figure}
\epsfxsize=3in
\begin{center}
\leavevmode
\epsffile{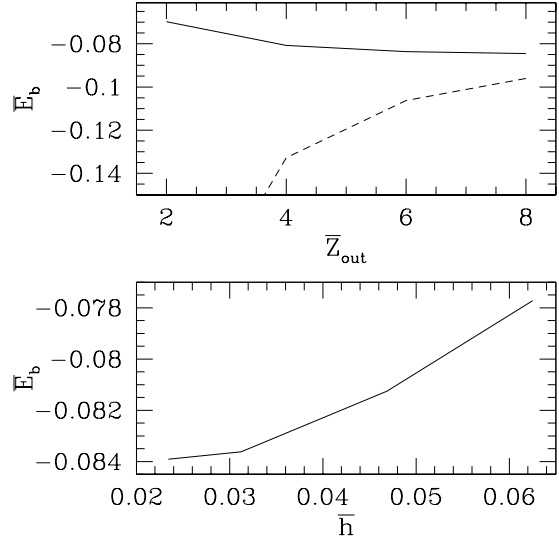}
\end{center}
\caption{The binding energy $\bar E_b$ for $\bar C = 2.5$ and $\bar J
= 3.0$ for different locations of the outer boundary $\bar Z_{\rm
out}$ (at constant grid resolution $\bar h = 0.03125$, top panel) and
different grid resolutions $\bar h$ (at constant outer boundary $\bar
Z_{\rm out} = 6$, bottom panel).  The dashed line only includes
contributions to the ADM mass from inside the computational grid, and
the solid line denotes the corrected value (see text). }
\end{figure}

The location of the outer boundary of the computational grid affects
the results through the Robin boundary condition on $u$, which is
correct only asymptotically, and the energy integral~(\ref{E}), which
should extend over all space.  The latter effect can be improved by
extending the energy integral beyond the numerical grid, where
$\alpha$ and $\beta$ can be evaluated analytically, and where $u$ can
be estimated from its value on the surface of the computational grid
and its $1/r$ falloff.  In the top panel of Fig.~1, we show the
binding energy for different locations of the outer boundary $\bar
Z_{\rm out} \equiv Z_{\rm out}/m$ for a typical configuration of
interest ($\bar C = 2.5$, $\bar J = 3.0$) with a fixed grid resolution
$\bar h \equiv h/m = 0.03125$.  For all calculations presented in this
paper we use $\bar X_{\rm out} = \bar Y_{\rm out} = \bar Z_{\rm
out}/2$.  For the dashed line, only contributions to the ADM mass from
inside the computational grid have been taken into account, and for
the solid line we have expanded the volume for the energy integral by
a factor of six in each dimension.  Obviously, the latter converges
much more rapidly and yields a more accurate value for all locations
of the outer boundary.

The resolution of the individual black holes affects the accuracy with
which their apparent horizons can be located, and hence the accuracy
of their masses $\bar M$.  The effect of the resolution on the
binding energy is demonstrated in the bottom panel of Fig.~1, where we
show $\bar E_b$ for different grid resolutions for the same
configuration, this time with the outer boundary fixed at $\bar Z_{\rm
out} = 6$.

From Fig.~1, we find that the binding energy $\bar E_b$ can be
determined to within at most a few percent error for $\bar Z_{\rm out}
= 6$ and $\bar h = 0.03125$, corresponding to a numerical grid of size
$96 \times 96 \times 192$.  The iteration to construct one model then
takes approximately 3 CPU hours on the NCSA Origin2000, which makes an
extensive survey of parameter space affordable.  We use these grid
specifications for all results presented in the following section.


\section{RESULTS}

\begin{figure}
\epsfxsize=3in
\begin{center}
\leavevmode
\epsffile{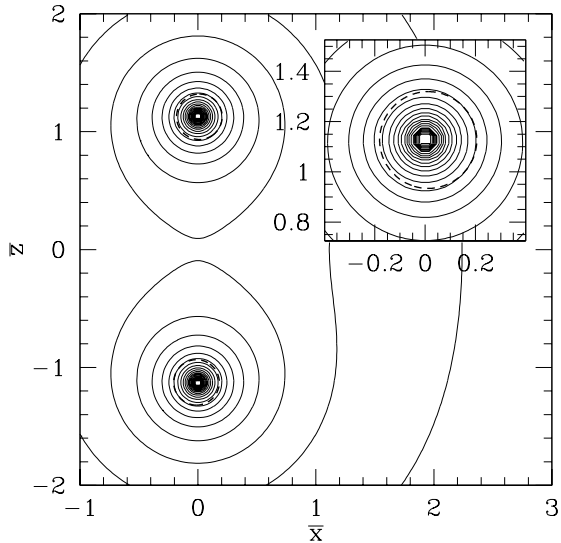}
\end{center}
\caption{Contours of the conformal factor $\psi$ for a configuration
close to the innermost stable circular orbit ($\bar C = 2.25$ and
$\bar J = 2.95$).  The contours (solid lines) logarithmically span the
interval $\psi = 1$ and $\psi = 9.2$.  Note that the apparent
horizons, marked by the thick dashed lines, are not concentric with
the contours of the conformal factor.  Instead, they are dragged along
by the black holes and lag slightly behind in their
(counter-clockwise) orbit.}
\end{figure}

In Fig.~2, we show contour plots of the conformal factor $\psi$ for a
configuration close to the ISCO ($\bar C = 2.25$ and $\bar J = 2.95$).
The apparent horizons, marked by the thick dashed lines, are dragged
along by the black holes and lag slightly behind in their
counter-clockwise orbit.  This effect has been discussed for single
boosted black holes in~\cite{cy90}.  Note that we compute the proper
separation $\bar l \equiv l/m$ between the horizons along the line
connecting the centers of the apparent horizons.  This is a
coordinate-dependent quantity, but a very good approximation to the
(coordinate-independent) shortest proper separation between the
horizons.

We now construct sequences of constant angular momentum for various
values of $\bar J$, and plot the effective potential $\bar E_b$ along
these sequences as a function of $\bar l$ in Fig.~3.  A minimum in the
effective potential corresponds to a stable quasicircular orbit.  The
bold line connecting these minima in Fig.~3 represents a sequence of
quasicircular orbits.  This sequence terminates at the ISCO, where the
adiabatic, quasicircular inspiral of the two black holes is expected
to change into a rapid plunge and merger~\cite{footnote0}.  Since this
transition leaves a characteristic signature in the gravitational wave
signal, the knowledge of the location and frequency of the ISCO are of
great importance for future observations with the new generation of
gravitational wave detectors.

\begin{figure}
\epsfxsize=3in
\begin{center}
\leavevmode
\epsffile{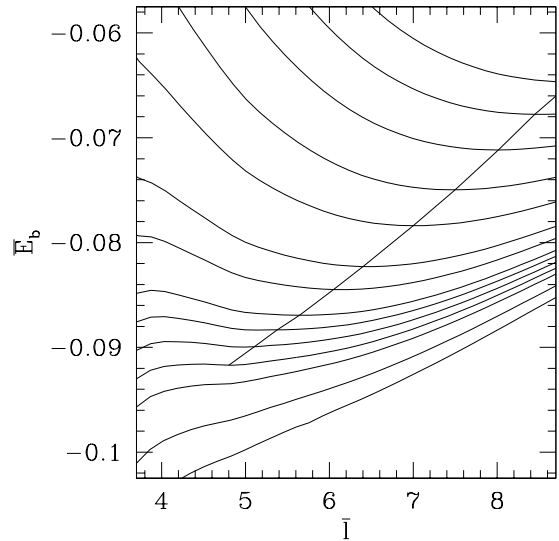}
\end{center}
\caption{The effective potential $\bar E_b$ as a function of proper
separation $\bar l$ for the following values of the angular momentum
$\bar J$: 2.9, 2.92, 2.94, 2.95, 2.96, 2.97, 2.98, 3.00, 3.02, 3.06,
3.10, 3.15, 3.20, 3.25 (from bottom to top).  Quasicircular orbits
correspond to minima in the effective potential.  The bold line
connects these minima and represents a sequence of quasicircular
orbits.  This sequence terminates at the innermost stable circular
orbit.}
\end{figure}

\begin{table}
\begin{center}
\begin{tabular}{llllll}
$\bar l$ & $\bar E_b$ & $ \bar J$&$\bar \Omega$ & $P/a$ & $C/a$ \\
\tableline 
 4.8	& -0.092	& 2.95	& 0.18	& 1.7	& 5.9\\
 4.880	& -0.09030	& 2.976 & 0.172	& 1.685 & 5.91\\
\end{tabular}
\end{center}
\caption{Comparison of our results for the innermost stable circular
orbit (top line) with those of Cook~\protect\cite{c94} (bottom line).}
\end{table}

In Table~I, we list our results for the physical parameters of the
ISCO (top line) and compare with the results of Cook~\cite{c94} (bottom
line).  We tabulate the proper separation between the horizons $\bar
l$, the binding energy $\bar E_b$, the angular momentum $\bar J$, and
the angular velocity $\bar \Omega$ as well as the the linear momentum
of each black hole $P/a$ and the coordinate separation $C/a$, where
$a$ is the (coordinate) radius of the black holes.  The latter is not
well-defined in our calculation, but since the black holes are nearly
spherical it is very reasonable to estimate an average radius from the
$\ell = 0$ monopole term in the multipole expansion for the horizon.

We conclude that all quantities agree fairly well with those of
Cook~\cite{c94} within our estimated numerical error of a few percent.
As the most significant deviation, we find that in our calculation the
binary is slightly more tightly bound at the ISCO, and correspondingly
has a slightly larger angular velocity.  However, even these
quantities differ by less than $\sim 5\%$, which may be caused by
numerical effects.  We conclude that the choice of the underlying
manifold structure has a very small effect on the location of the
ISCO.

\begin{table}
\begin{center}
\begin{tabular}{llll}
Reference 		& $\bar E_b$ 	& $ \bar J$	&$\bar \Omega$ \\
\tableline 
Schwarzschild 		& -0.0572 	& 3.464		& 0.068 \\ 
CE~\cite{ce77}  	& -0.1		& 3.3 		& ---	\\
BD~\cite{bd92}  	& -0.65		& 0.85 		& 2	\\
KWW~\cite{kww92}	& -0.0378	& 3.83		& 0.0605\\
Cook~\cite{c94} 	& -0.09030	& 2.976 	& 0.172	\\
BCSST~\cite{betal98}	& -0.048	& 3.9		& 0.06	\\
DIS~\cite{dis98}	& -0.0653	& ---		& 0.0885\\
BD~\cite{bd99} 		& -0.06005	& 3.40		& 0.0734\\
This work		& -0.092	& 2.95		& 0.18  \\
\end{tabular}
\end{center}
\caption{Comparison of the binding energy $\bar E_b$, the angular
momentum $\bar J$ and the angular velocity $\bar \Omega$ for the
innermost stable circular orbit from various different calculations: a
test particle orbiting a Schwarzschild black hole, Clark and Eardley
(CE)~\protect\cite{ce77}, Blackburn and Detweiler
(BD)~\protect\cite{bd92}, Kidder, Will and Wiseman
(KWW)~\protect\cite{kww92}, Cook~\protect\cite{c94}, Baumgarte {\it
et.al.} (BCSST)~\protect\cite{betal98}, Damour, Iyer and Sathyaprakash
(DIS)~\protect\cite{dis98}, Buonanno and Damour
(BD)~\protect\cite{bd99}, and the results from this paper.  The
results of Baumgarte {\it et.al.}~\protect\cite{betal98} are for an
$n=1$ polytrope binary of compaction $M/R = 0.2$.  Naively
extrapolating to $M/R = 0.5$ yields values very close to our results
for binary black holes.}
\end{table}

In Table~II we compare the binding energy, the angular momentum and
the angular velocity at the ISCO from various different calculations.
For a test particle orbiting a Schwarzschild black hole, the ISCO can
be located analytically, which yields $\bar E_b = \sqrt{8/9} - 1 \sim
-0.0572$, $\bar J = 2 \sqrt{3} \sim 3.464$, and $\bar \Omega =
1/6^{3/2} \sim 0.0680$.  Clark and Eardley~\cite{ce77} adopted a first
order post-Newtonian argument to approximately estimate the location
of the ISCO.  We list their values for nonrotating neutron stars.
Blackburn and Detweiler~\cite{bd92} adopted a variational principle
and assumed a periodic solution to Einstein's equations.  At the ISCO,
for which the approximations of this approach fail, they find
extremely tightly bound binaries.  Kidder, Will and
Wiseman~\cite{kww92} adopted a second order post-Newtonian
approximation together with a ``hybrid'' approach and found an
extremely weakly bound ISCO.  However, several authors have cast doubt
on the robustness and consistency of the hybrid
approach~\cite{ws93,dis98}.  Baumgarte {\it et.al.}~\cite{betal98}
constructed fully relativistic models of corotating binary neutron
star in quasiequilibrium, albeit assuming conformal flatness, and
found that the ISCO depends on the compaction of the neutron stars.
In Table~I we list their results for $n=1$ polytropes of compaction
$M/R = 0.2$, where $M$ and $R$ are the mass-energy and areal radius
which the stars would have in isolation.  Naively extrapolating their
results to $M/R = 0.5$ yields values which are very similar to our
result for binary black holes.  Damour, Iyer and
Sathyaprakash~\cite{dis98} combined a second order post-Newtonian
approximation with a Pad\'{e} approximation, which yields a slightly
more tightly bound ISCO than that for a test-particle orbiting a
Schwarzschild black hole.  A similar result is found by Buonanno and
Damour, who combine a second order post-Newtonian approximation with
an effective-one-body method~\cite{bd99,bd00}.


\section{SUMMARY AND DISCUSSION}

Since an accurate knowledge of the ISCO is very important for possible
future gravitational wave observations, it is very unsettling that
different approaches to computing the ISCO lead to very different
results (compare Table~II).  One of these approaches, namely constructing
binary black hole solutions to the constraint equations of general
relativity, involves {\em choosing} the background geometry and
topology, and it would be very desirable to know how much the results
depend on these choices.

In this paper, we introduce a new method to construct solutions to the
constraint equations of general relativity describing binary black
holes in quasicircular orbit.  We combine the approaches of
Cook~\cite{c94} and Brandt and Br\"ugmann~\cite{bb97} to construct
binary black holes in a three-sheeted manifold structure, as opposed
to the two-sheeted topology in the conformal-imaging approach adopted
by Cook~\cite{c94}.  We locate the ISCO and find that its physical
parameters are very similar to those found by Cook~\cite{c94}.  Our
results confirm those earlier results and imply that the underlying
manifold structure only has a very small effect on the ISCO.  The
latter is perhaps not entirely surprising, since it reflects the fact
that the strength of the imaged poles in the conformal-imaging
approach is smaller than the strength of the poles
themselves~\cite{m63}.

Our new approach is considerably simpler than the conformal-imaging
approach of Cook~\cite{c94}.  The analytic solution to the momentum
constraint simplifies because no inversion-symmetric solutions have to
be constructed, and the numerical solution to the Hamiltonian
constraint simplifies because we can adopt the ``puncture'' method of
Brandt and Br\"ugmann~\cite{bb97}.  In particular, we can solve the
Hamiltonian constraint in cartesian coordinates on $R^3$ without
having to impose interior boundary conditions.  One disadvantage
of our approach is that the apparent horizons have to be located
numerically, which we do with the algorithm developed in~\cite{betal96}.

In this paper, we follow Cook~\cite{c94} and choose a conformally flat
background metric.  Accordingly, we cannot address the dependence of
the ISCO on the choice of the background geometry.  However, since our
new method is significantly simpler than the conformal-imaging
approach, it may provide a useful framework to relax the assumption of
conformal flatness and to construct binary black holes in
quasicircular orbit for more general background geometries.


\acknowledgments

It is a pleasure to thank G. B. Cook, S. L. Shapiro and M. Shibata for
various very useful conversations.  The author gratefully acknowledges
the hospitality at the Institute for Theoretical Physics, Santa
Barbara, where this research was initiated during the mini-program on
``Colliding Black Holes''.  Calculations were performed on SGI CRAY
Origin2000 computer systems at the National Center for Supercomputing
Applications, University of Illinois at Urbana-Champaign.  This work
was supported by NSF Grants PHY 99-02833 at Illinois and PHY 94-07194
at Santa Barbara.



\begin{references}

\bibitem{itp00} See, {\it e.g.}, the talks presented at the ITP 
	Miniprogram {\em Colliding Black Holes: Mathematical Issues 
	in Numerical Relativity} ({\tt doug-pc.itp.ucsb.edu/online/numrel00/}).

\bibitem{footnote0} Note that the transition from adiabatic inspiral
	to dynamical plunge at the innermost stable circular orbit
	may be fairly gradual~\cite{bd00,ot00}.

\bibitem{bd00} A. Buananno and T. Damour, submitted (2000), also
	gr-qc/0001013.

\bibitem{ot00} A. Ori and K. S. Thorne, submitted (2000), also 
	gr-qc/0003032.

\bibitem{ce77} J. P. A Clark and D. M. Eardley, Astrophys. J. {\bf 215},
	311 (1977).

\bibitem{bd92} J. K. Blackburn and S. Detweiler, Phys. Rev. D {\bf 46},
	2318 (1992).

\bibitem{kww92} L. E. Kidder, C. M. Will and A. G. Wiseman, Class.
	Quantum Grav. {\bf 9}, L125 (1992);  
	Phys. Rev. D {\bf 47}, 3281 (1993).

\bibitem{dis98} T. Damour, B. R. Iyer and B. S Sathyaprakash, 
	Phys. Rev. D {\bf 57}, 885 (1998).

\bibitem{bd99} A. Buananno and T. Damour, Phys. Rev. D {\bf 59}, 084006 (1999).

\bibitem{c94} G. B. Cook, Phys. Rev. D {\bf 50}, 5025 (1994).

\bibitem{betal98} T. W. Baumgarte, G. B. Cook, M. A. Scheel, 
	S. L. Shapiro, and S. A. Teukolsky,
	Phys. Rev. Lett. {\bf 79}, 1182 (1997);
	Phys. Rev. D {\bf 57}, 7292 (1998).

\bibitem{bb97} S. Brandt and B. Br\"ugmann, 
	Phys. Rev. Lett. {\bf 78}, 3606 (1997).

\bibitem{adm62} R. Arnowitt, S. Deser and C. W. Misner, in 
        {\em Gravitation: An Introduction to Current Research}, 
        edited by L. Witten (Wiley, New York, 1962).

\bibitem{y71} J. W. York, Jr., Phys. Rev. Lett. {\bf 26}, 1656 (1971).

\bibitem{y79} J. W. York, Jr., in {\em Sources of Gravitational Radiation},
	edited by L. L. Smarr (Cambridge University Press, Cambridge,
	England, 1979).

\bibitem{c91} G. B. Cook, Phys. Rev. D {\bf 44}, 2983 (1991).

\bibitem{cetal93} G. B. Cook, M. W. Choptuik, M. R. Dubal, S. Klasky,
	R. A. Matzner and S. R. Oliveira, Phys. Rev. D {\bf 47}, 1471 (1993).

\bibitem{er35} A. Einstein and N. Rosen, Phys. Rev. {\bf 48}, 73 (1935).

\bibitem{l63} R. W. Linquist, J. Math. Phys. {\bf 4}, 938 (1963).

\bibitem{m63} C. W. Misner, Ann. Phys. (N.Y.) {\bf 24}, 102 (1963).

\bibitem{ksy83} A. D. Kulkarni, L. C. Shepley and J. W. York, Jr.,
	Phys. Lett. {\bf 96A}, 228 (1983).

\bibitem{c71} A. \v{C}ade\v{z}, Ph.D. dissertation, University of 
	North Carolina (1971); see also Appendix C in~\cite{c91}.

\bibitem{footnote2} Note that these integrals are in terms of cartesian
	coordinates.

\bibitem{by80} J. Bowen and J. W. York, Jr., 
	Phys. Rev. D {\bf 21}, 2047 (1980).

\bibitem{b79} J. Bowen, Gen. Relativ. Gravit. {\bf 11}, 227 (1979).

\bibitem{y89} J. W. York, Jr., in {\em Frontiers in Numerical Relativity},
	edited by C. R. Evans, L. S. Finn, and D. W. Hobbill (Cambridge
	University Press, Cambridge, 1989).

\bibitem{c70} C. Christodoulou, Phys. Rev. Lett. {\bf 25}, 1596 (1970).

\bibitem{footnote3} To illustrate this approach, consider two Newtonian
	point-masses of mass $M$ at a separation $l$, for which 
	$E_b = - M^2/l + J^2/l^2M$.  Minimizing $E_b$ for constant
	$M$ and $J$ yields $J^2 = M^3 l/2$ for circular orbits.  The
	orbital angular velocity can then be found from $\Omega = 
	\partial E_b/\partial J = 2J/l^2M$, where the derivative is taken
	for constant $M$ and $l$.  Combining the two results recovers,
	as expected, the Kepler law $\Omega = \sqrt{2M/l^3}$.  
	
\bibitem{dagh} M. Parashar and J. C. Brown, in {\em Proceedings of 
        the International Conference for High Performance Computing}, 
        edited by S. Sahni, V. K. Prasanna and V. P. Bhatkar
        (Tata McGraw-Hill, New York, 1995), also 
        {\tt www.caip.rutgers.edu/$\sim$parashar/DAGH/}.

\bibitem{petsc} S. Balay, W. D. Gropp, L. C. McInnes, and B. F. Smith,
	in {\em Modern Software Tools in Scientific 
        Computing}, edited by E. Arge, A. M. Bruaset and H. P. Langtangen,
        p.~163, Birkhauser Press, 1997; see also the PETSc home page
        {\tt http://www.mcs.anl.gov/petsc}.

\bibitem{b00} B. Br\"ugmann, private communication.

\bibitem{betal96} T. W. Baumgarte, G. B. Cook, M. A. Scheel, 
	S. L. Shapiro, and S. A. Teukolsky,
	Phys. Rev. D {\bf 54}, 4849 (1996).

\bibitem{footnote4} Alternatively, all quantities could be rescaled
	with respect to the {\em numerical} values of $M$.  In this
	case, no iteration is needed to achieve $\bar M = M_{\rm
	num}/M = 1$, since this obviously holds identically, but
	instead an iteration is necessary to find a desired value of
	$\bar J$ (see~\cite{c94}).

%

\bibitem{cy90} G. B. Cook and J. W. York, Jr., 
	Phys. Rev. D {\bf 41}, 1077 (1990).

\bibitem{ws93} N. Wex and G. Sch\"afer, Class. Quantum Grav. {\bf 10},
	2729 (1993); also G. Sch\"afer and N. Wex, in {\em XIIIth Moriond 
	Workshop: Perspectives in Neutrinos, Atomic Physics and Gravitation},
	edited by J. Tr\^{a}n Thanh V\^{a}n, T. Damour, E. Hinds and
	J. Wilkerson (Editions Fronti\`{e}res, Gif-sur-Yvette, 1993), p.~513.


\end{references}
\end{document}